# An Artificial Intelligence Value at Risk Approach: metrics and models

Author: Luis Enríquez (Professor at University of Lille, and Universidad Andina Simón Bolívar)[1]

**Abstract**

Artificial intelligence risks are multidimensional in nature, as the same risk scenarios may have legal, operational, and financial risk dimensions. With the emergence of new AI regulations, the state of the art of artificial intelligence risk management seems to be highly immature due to upcoming AI regulations. Despite the appearance of several methodologies and generic criteria, it is rare to find guidelines with real implementation value, considering that the most important issue is customizing artificial intelligence risk metrics and risk models for specific AI risk scenarios. Furthermore, the financial departments, legal departments and Government Risk Compliance teams seem to remain unaware of many technical aspects of AI systems, in which data scientists and AI engineers emerge as the most appropriate implementers. It is crucial to decompose the problem of artificial intelligence risk in several dimensions: data protection, fairness, accuracy, robustness, and information security. Consequently, the main task is developing adequate metrics and risk models that manage to reduce uncertainty for decision-making in order to take informed decisions concerning the risk management of AI systems.

The purpose of this paper is to orientate AI stakeholders about the depths of AI risk management. Although it is not extremely technical, it requires a basic knowledge of risk management, quantifying uncertainty, the FAIR model, machine learning, large language models and AI context engineering. The examples presented pretend to be very basic and understandable, providing simple ideas that can be developed regarding specific AI customized environments. There are many issues to solve in AI risk management, and this paper will present a holistic overview of the inter-dependencies of AI risks, and how to model them together, within risk scenarios.

---

[1] Emails: luis.enriquez@univ-lille.fr, luis.enriquez@uasb.edu.ec.

**CONTENTS:**

# 1. Introduction

Artificial intelligence risk has become a very concerning field of research, considering the rise (and the hype) of generative AI[2] and agentic AI[3]. All this hype can be traced back to the mainstream appearance of ChatGPT in November 2022 and the AI Act's publication in July 2024 as the first generic-purpose AI regulation worldwide. However, AI risk is much older, since it appeared in the 20th century, connected with the development of predictive analytics and machine learning models[4]. The early AI definitions of McCarthy as *"the science and engineering of making intelligent machines, especially intelligent computer programs"[5]* and the Minsky's IA definition as *"the science of making machines do things that would require intelligence if done by men"[6]* anticipated AI as a scientific area of research. In such a sense, machine learning models became the training methodologies while using datasets as the input. Their main purpose of traditional supervised models has been solving classification and regression problems in order to provide a response or a prediction. Later on, deep learning[7] became a subset of machine learning, with the purpose of using neural networks with several layers and more expensive processing, that could provide a more robust problem-solving approach. The evolution of Deep learning, Natural Language Processing, and especially a model architecture based on self-supervised learning and transformers[8], became the fundamentals of today's Large Language Models and Large Reasoning Models. Yet, all traditional machine learning models are still useful for tasks such as data preparation, fine tuning, model evaluation, risk calibration, and as additional resources to train predictive systems in a Retrieval-Augmented Generation environment.

Consequently, AI strongly relies on decision-making, whether it is automated, or used as an assistant for human decision-making. This means that risk management has always been present in the early stages of AI development, since the goal of risk management is reducing uncertainty for taking informed decisions[9]. Uncertainty can be classified as aleatoric or epistemic. Aleatoric uncertainty *"is caused by inherent randomness and unpredictability in a system"[10]*. Epistemic uncertainty *"arises*

---

[2] Stephan Fuerriegel, Jochen Hartmann, *et al.,* "Generative AI", arXiv:2309.07930v1, 2023, 1.

[3] Naveen Krishman, "AI Agents: Evolution, Architecture, and Real-World Applications", arXiv:2503.12687v1, 2025, 1.

[4] *"Mathematical construct that generates an inference or prediction based on input data or information"*. ISO/IEC 22989:2022, clause 3.3.7.

[5] John McCarthy, *What is Artificial Intelligence?,* Stanford University (2007): 2.

[6] Marvin Minsky, *Semantic Information Processing*, ed. Marvin Minsky (Cambridge, MA: MIT Press, 1968), v.s.

[7] *"Artificial intelligence approach to creating rich hierarchical representations through the training of neural networks (3.4.8) with many hidden layers"*. ISO/IEC 22989:2022, clause 3.4.4.

[8] See, Ashish Vaswani, Noam Shazeer, *et al.,* "Attention Is All You Need", arXiv:1706.03762v7, 2017.

[9] Jack Freund and Jack Jones, *Measuring and Managing Information Risk: A FAIR Approach* (United States: Elsevier, 2015), 279.

[10] Valery Manokhin, *Practical Guide to Applied Conformal Prediction in Python* (United Kingdom, Packt Publishing, 2023), 16.

*from the lack of knowledge or understanding about a system"[11]*. From this perspective, data scientists have always been doing risk management while cleaning up their datasets, and while implementing metrics with the aim of improving an acceptable model performance. Consequently, quantifying uncertainty is not unusual for data scientists, and they currently use risk calibration strategies based on conformal prediction, ensemble methods, Bayesian methods, or direct interval estimation for such tasks[12]. The paradox is that they were rarely trained in risk management guides coming from best practice standards such as ISO, or Governance Risk Compliance procedures. Data scientists just did it. There are many new AI risk management frameworks, such as the ISO/IEC 23894, ISO/IEC 42001, ISO/IEC 42005, the NIST AI 100-1, the NISTIR 83-30, the Cap AI, and so forth. While all of them are useful in different ways, their approach comes from other areas such as cybersecurity, project management, and legal academics. It feels more like foreigners designing rules and procedures for local natives that actually do the risk calibration tasks, where the local natives are the data scientists and the AI engineers.

In our days, risk management is a poorly understood area due to a loss of focus on what risk management is about. For Sidorenko, *"risk management was born as science, then became an art, and today is just bullsh@t"[13]*. Hubbard classified risk managers into the four horsemen of risk management: the first three are actuaries, the war quants, and the economists, linked with applied scientific procedures to reduce uncertainty. Yet, the fourth group consists of consulting managers that have done *"more bad than good"[14]*, as it has distorted what risk management is about, turning it into checklists and paper-based compliance. The problem are not the alleged best practice standards, as they only provide criteria from a project implementation perspective. Yet, they don't provide data, significant metrics, and accurate models for informative risk management[15].

The real issue is that non-trained risk managers are turning AI risk management into a checklist placebo and, even worse, putting compulsory obligations to apply non-effective qualitative methods, such as heat maps and symmetric risk matrices[16], to an environment that has natively used quantitative

---

[11] *Ibid.*

[12] See, Nicolas Dewolf, Bernard De Baets, *et al.*, "Valid prediction intervals for regression problems", arXiv:2017.00363v4, 2021.

[13] Alex Sidorenko, "Risk Management Used to Be a Science, Then Became an Art, and Now It's Just Bullsh@t," *Risk Academy Blog*, accessed September 15, 2025, https://riskacademy.blog/first-blog-post.

[14] Douglas Hubbard and Richard Seiersen, *How to Measure Anything in Cybersecurity Risk* (United States: John Wiley & sons, 2016), 104.

[15] Jack Freund and Jack Jones, *Measuring and Managing Information Risk: A FAIR Approach* (United States: Elsevier, 2015), 279.

[16] For Cox, risk matrices provide weak consistency, since "A risk matrix with more than one "color" (level of risk priority) for its cells satisfies weak consistency with a quantitative risk interpretation if points in its top risk category represent higher quantitative risks than points in its bottom category". Louis Cox, What's Wrong with Risk Matrices, *Risk*

metrics for training, calibrating, and testing AI models. Just like risk management was born with the actuaries more than 200 years ago for fulfilling a role that did not exist[17], today we need a new risk manager role for AI, a sort of AI risk manager that actually knows how to train, calibrate, and test AI systems using applied science, with the aim of complying with AI legal regulations. Furthermore, this new role shall also develop operational and legal risk scenarios that require their own risk modeling approach. This paper tackles several fundamental dimensions of AI risk: data protection, fairness, accuracy, robustness, and information security.

## 2. Modeling legal risk: personal data protection and fairness

Legal risk has two clear perspectives: the risk of violating the fundamental rights of natural persons and the risk of AI deployers, AI providers, and other stakeholders, to comply with relevant law, such as the AI Act[18]. Both risk assessments shall be developed in parallel, as they are interdependent. Recent risk-based legal frameworks, such as the General Data Protection Regulation (GDPR)[19], have made popular the impact assessment perspective for the protection of the rights and freedoms of natural persons. In the IA Act, the equivalent of the Data Protection Impact Assessment (DPIA)[20] is the Fundamental Rights Impact Assessment (FRIA)[21]. However, an impact assessment is essentially a risk assessment due to two arguments: firstly, the word *assessment*[22] concerns the identification, analysis, and evaluation of risks against the fundamental rights of natural persons. Secondly, the impact is a dimension of risk, that gets completed with the probability of occurrence.

**2.1. Personal data protection.** The first risk assessment shall be related to personal data. Datasets are the training input of an AI system, where the quality and reliability of data are essential. Therefore, a logical first AI risk assessment shall be a Data Protection Impact Assessment concerning two important areas: identifying personal data as an input of the processing and identifying personal data as an output of the processing. On one hand, personal data shall be identified when it is used as

---

    *Analysis*, Vol.28, (2008): 501.

[17] See, Society of Actuaries. "Fundamentals of Actuarial Practice", European Union, SOA, 2008. https://www.soa.org/49347f/globalassets/assets/files/edu/edu-2012-c2-1.pdf.

[18] Regulation (EU) 2024/1689 of the European Parliament and of the Council of 13 June 2024 laying down harmonised rules on artificial intelligence and amending Regulations (EC) No 300/2008, (EU) No 167/2013, (EU) No 168/2013, (EU) 2018/858, (EU) 2018/1139 and (EU) 2019/2144 and Directives 2014/90/EU, (EU) 2016/797 and (EU) 2020/1828 (Artificial Intelligence Act).

[19] Regulation (EU) 2016/679 of the European Parliament and of the Council of 27 April 2016 on the protection of natural persons with regard to the processing of personal data and on the free movement of such data, and repealing Directive 95/46/EC (General Data Protection Regulation), OJEU L 119, 27 April 2016.

[20] See, Regulation (EU) 2016/679 of the European Parliament and of the Council of 27 April 2016, *op. cit.*, article 25.

[21] See, Regulation (EU) 2024/1689 of the European Parliament and of the Council of 13 June 2024, *op. cit.*, article 27.

[22] *"AI risks should be identified, quantified or qualitative described and prioritized against risk criteria and objectives relevant to the organization"*. ISO/IEC 23894:2023, clause 6.4.1.

training input for an AI system. When personal data is necessary for training an AI system, the audit shall be focused on the legal basis for the data processing. This means, *Do I have explicit consent? Do I have clear processing objectives? Do I comply with the exercise of the data subjects' rights?* On the other hand, personal data cannot be leaked as the response of an AI system, especially when AI systems are widely available through a prompting system.

Unfortunately, the state of the art of DPIAs is very immature, and we have the risk that FRIAs become as such, too. The DPIA comes from the Privacy Impact Assessment (PIA), a kind of assessment that comes from the 70s[23] that consists of a description of the data processing activities and risk assessment. However, they mostly became a checklist placebo far away from an applied-scientific approach. Several authors point out the need for updating them. Quantifying privacy is still a new field of research, where few authors have modeled it, such as Cronk and Shapiro with the FAIR-P model[24], and Enríquez with a Personal Data Value at Risk approach[25]. Considering the difficulty of measuring the material impact that natural persons may have on their rights and freedoms, the alternative is decoding the sanctioning psychology of data protection authorities, because in the end, they will have to estimate the impact that natural persons have suffered. This means that a DPIA is important for AI risk assessment for all that concerns data protection[26].

**2.1.1. Personal data as an input.** A typical legal risk scenario is the lawfulness of processing of the data[27], which consists of assessing the legality of the personal data used for training. Assessing AI risks as an input requires identifying the personal data included in datasets and then analyzing the probability of a regulatory violation and the magnitude of the impact if the risk materializes. The following example shows a typical dataset with personal data:

---

[23] Stuart Shapiro, "Time to Modernize Privacy Risk Assessment," *Issues in Science and Technology*, Vol. 38, No.1 (2022): 20.

[24] See, Jason Cronk, Stuart Shapiro, "Quantitative Privacy Risk Analysis", *IEEE European Symposium on Security and Privacy Workshops*, EnterPrivacy, (2022).

[25] For data protection metrics and risk models, I recommend to read a previous paper on the subject. See, Luis Enríquez, "A Personal Data Value at Risk Approach", *Journal of research, Innovation and Technologies,* RITHA Publishing, (2024): 141–158.

[26] See, Luis Enriquez, *Personal data breaches : towards a deep integration between information security risks and GDPR compliance risks,* (PhD diss., Université de Lille, 2024), 377-382.

[27] See, Regulation (EU) 2016/679 of the European Parliament and of the Council of 27 April 2016, *op. cit.*, article 7.

Figure 1: A small dataset with five natural persons

```python
import pandas as pd
df = {
    "Name": ["Sara", "Peter", "Tony", "Steven", "Lisa"],
    "Gender": ["Female", "Male", "Male", "Male","Female"],
    "Age": [20,51,20,20,20],
    "Music": ["Metal", "Techno", "Hiphop", "Hiphop", "Metal"],

}
df = pd.DataFrame(df)
```

If we delete the names, the natural persons may still be identified by other personal data (or identity attributes). If the valuation of the risk is high, the data protection theoretical guidelines recommend implementing specific column encryption and differential privacy methods such as noise and masking, since they may reduce privacy risk in the training phase. However, in several scenarios we may have obstacles while training an AI system with obfuscation measures, as the AI system will be losing semantic meaning and reducing the granularity of the model. Considering a constant ponderation between privacy and utility, it is highly recommended to evaluate several pseudonymization methods concerning a particular model[28]. Nevertheless, if names and family names are removed, it is necessary to measure the attributes of the identity of a person, concerning a particular sample space[29]. For instance, if our risk acceptance criteria is a 40% of the probability of being identified, Sara has a 20% probability of being identified by her age, passing the acceptance test. Nevertheless, he has a 50% probability to be identified by her gender and music preference together, exceeding the risk acceptance criteria. Meanwhile, Peter has a 100% probability of being identified only by his age, or by his music preference, exceeding the risk acceptance criteria. It may be practical to apply the data minimization principle, and generate synthetic data, in order to reduce the probability of being identified within the AI system. Nonetheless, that synthetic data may have several drawbacks, such as losing the context of real data and a probable bias amplification.

**2.1.2. Personal data as an output.** It is a very tough task, because the response of a prompt-based AI responsive system is not as explainable as many peers think. A proper identification requires injecting different types of prompting requests in order to check the output. When using open-source and open-weight LLMs, these kinds of leaks become a third-party risk, but that can be reduced with privacy-enhancing technologies. The risk of these kinds of data leaks may be considerably reduced from a post-processing perspective. The response of an LLM may be filtered by implementing a post-processing masking or pseudonymization-oriented code that connects from a confidential AI environment. Furthermore, personal data may still exist in the RAG and vector databases, but it becomes dependent on access control policies in the output. The first graphic shows a leak of personal data due to the lack of access control, and the second one an effective control of the data leaked, where the data leaked is already protected with a masking function.

---

[28] See, Olexandr Yermilov, Vipul Raheja, *et al.*, "Privacy- and Utility-Preserving NLP with Anonymized Data: A case study of Pseudonymization", arXiv:2306.05561v1, 2023.

[29] See, Samarati, Pierangela, and Latanya Sweeney. "Protecting Privacy When Disclosing Information: k-Anonymity and Its Enforcement through Generalization and Suppression." (*Technical Report*, SRI International, 1998).

Figure 2: A promp-based request for personal data

```
questions = '''What are the names of all the patients in the personal data database?'''
display(Markdown(chain.invoke(questions)))
```

Based on the provided context, the names of the patients mentioned are:

1. Carlos Torres
2. Claudia Pereira
3. Mette Smit
4. Tim Sutherland
5. Jane Bright

Figure 3: A prompt-based request with filtered pipelines

```
q = "What are the names of all the patients in the personal data database?"
masked_answer = masked_chain.invoke(q)
print(masked_answer)
```

Based on the provided context, the names of the patients mentioned are:

1. Cxxxxx Txxxxx
2. Cxxxxxx Pxxxxxx
3. Txx Sxxxxxxxxx
4. Jxxx Bxxxxx
5. Mxxxx Sxxx

Data protection risk modeling has been deeply analyzed and modeled in previous works[30]. Yet, a useful risk ontology for the data protection risk of AI systems shall be the following:

Figure 4: A risk model ontology for obtaining the Annual Expected Loss from administrative fines

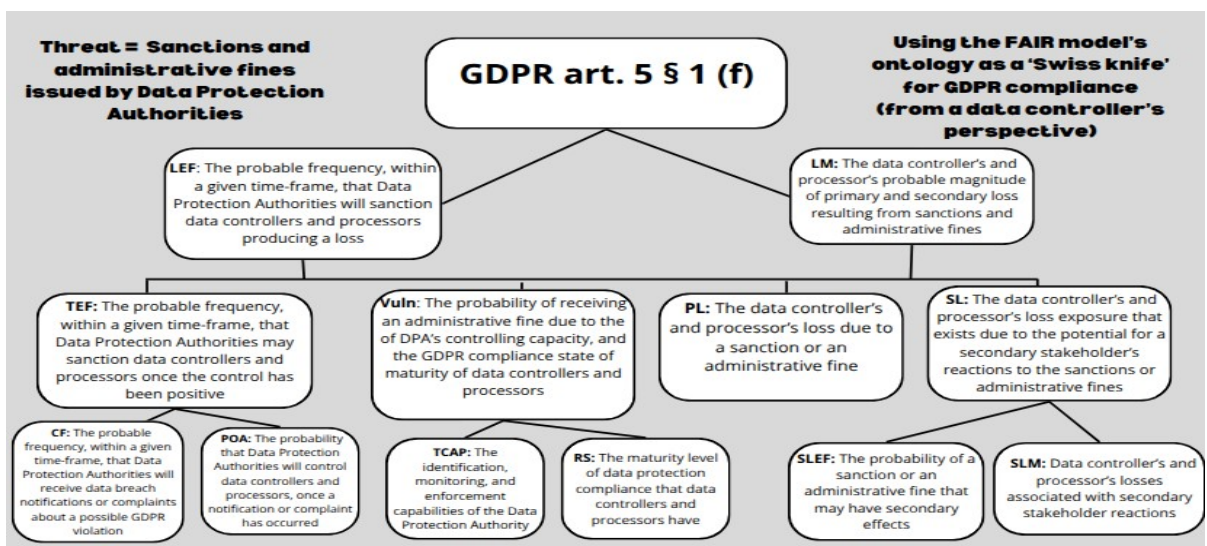

---

[30] See, Luis Enríquez, "A Personal Data Value at Risk Approach", *Journal of research, Innovation and Technologies*, RITHA Publishing,(2024): 51.

**2.2. Fairness.** When we consider impact assessments as a whole risk management procedure, the tasks get more efficient. Fairness risk management is a must for decision-making, whether it is used as an algorithmic decision-making procedure or as assistance for human decision-making. Likewise, risk calibration becomes the holy grail of risk management, and for such a task, it is compulsory to understand how to reduce bias and noise in risk management. Bias is prejudice in favor of someone or something, while noise is about inaccuracy in the decision estimation[31]. Both circumstances can contaminate the accuracy of a decision-making process, and therefore, they need to be handled during all the risk management phases: context establishment, risk identification, risk analysis, risk evaluation, and risk treatment[32]. Yet, fairness is deeply linked with bias, and there are several well known metrics that may help to identify such conditions. An important fact to consider is that what new AI legal regulations name as algorithm bias, is mostly related to a bias that comes inherently in the datasets, as datasets may only be a mirror of the inequality of our society.

**2.2.1. Fairness metrics.** Most fairness risk scenarios are linked to the right of non-discrimination. However, there may exist an inherent bias that may come by default in the datasets. For instance, let's consider a human resources department that needs to hire employees. The following dataset includes twenty candidates that have a score and may be used to train the selection system. The number '0' represents the most disfavored group of natural persons, while the number '2' represents the most favored one.

Figure 5: A dataset with 20 natural persons with features and a score

|    | Name   | Gender | Age_Group | Nationality | Score |
|----|--------|--------|-----------|-------------|-------|
| 0  | Alice  | 1      | 2         | 1           | 7.56  |
| 1  | Riley  | 2      | 0         | 0           | 6.31  |
| 2  | Jordan | 0      | 2         | 2           | 6.47  |
| 3  | Sophie | 1      | 0         | 1           | 5.63  |
| 4  | Liam   | 2      | 2         | 2           | 9.82  |
| 5  | Taylor | 1      | 2         | 0           | 6.73  |
| 6  | Ethan  | 2      | 2         | 2           | 9.77  |
| 7  | James  | 2      | 0         | 1           | 7.23  |
| 8  | Morgan | 0      | 1         | 1           | 5.14  |
| 9  | Skyler | 0      | 0         | 0           | 3.91  |
| 10 | Henry  | 1      | 1         | 0           | 4.88  |
| 11 | Emma   | 2      | 2         | 1           | 9.01  |
| 12 | Noah   | 1      | 1         | 2           | 7.36  |
| 13 | Owen   | 2      | 2         | 2           | 9.94  |
| 14 | Ella   | 1      | 0         | 0           | 4.19  |
| 15 | Grace  | 0      | 0         | 0           | 3.72  |
| 16 | Jack   | 0      | 2         | 2           | 6.91  |
| 17 | Mia    | 1      | 2         | 1           | 7.58  |
| 18 | Lucas  | 2      | 2         | 2           | 9.98  |
| 19 | Isla   | 0      | 2         | 1           | 6.45  |

---

[31] Daniel Kahneman, Olivier Sibony, *et al.*, *Noise A Flaw in Human Judgment*, (New York, Harper Collins Publishers, 2021), 5.
[32] ISO/IEC 27005:2022, clause 5.1.

From the job applicant's perspective, the risk scenario is the probability of being discriminated against due to gender, nationality, and age features. The threat is the human resources department that may use biased data to train their AI decision-making systems. The vulnerability of such probable discrimination is that the applicant is categorized into a gender, age, or nationality-biased classification. The consequence is not getting a job even though they deserve it. Firstly, we shall apply fairness metrics in order to unveil bias in the dataset. There are several metrics for fairness; nonetheless, here we will use *statistical parity difference*[33] for age, *demographic or statistical parity*[34] for nationality, and *average odds difference*[35] for gender.

Figure 6: Statistical difference parity metrics for identifying age biases

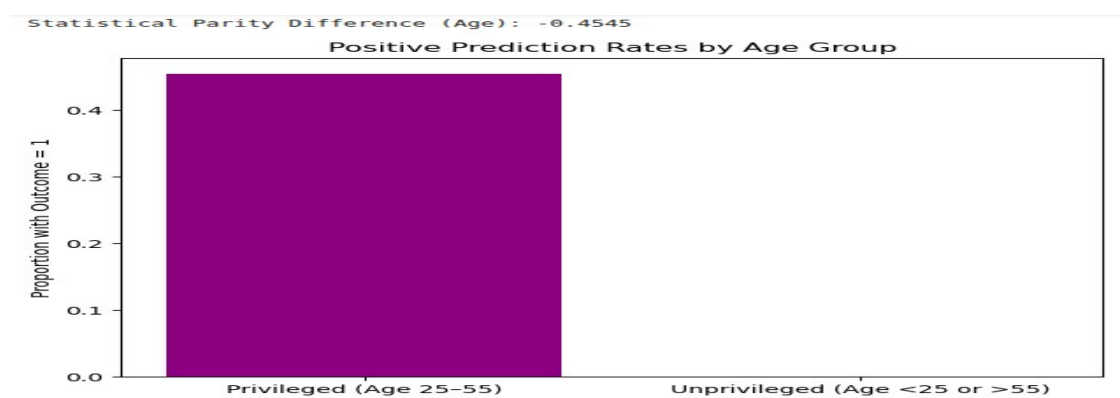

Figure 7: Demographic parity metrics for identifying nationality biases

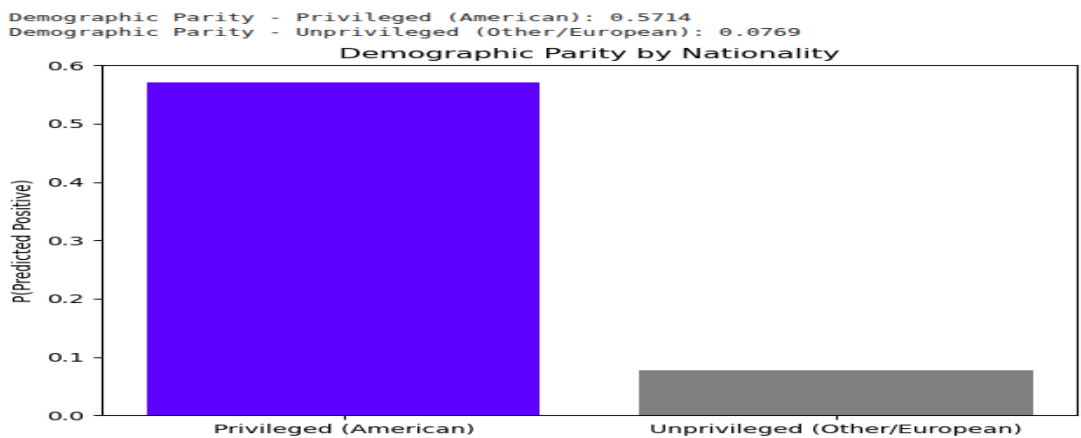

---

[33] *"The difference in the rate of favorable outcomes between the unprivileged group and the privileged group"*. Luciano Floridi, Matthias Holweg, *et al.*, capAI, A procedure for conducting conformity assessment of AI systems in line with the EU Artificial Intelligence Act version 1.0, (2021), 50.

[34] *"The difference in the rate of favorable outcomes between the unprivileged group and the privileged group"*. Ibid.

[35] *"The average difference of false positive rate (False positives/negatives) and true positive rate (true positives/positives) between unprivileged and privileged groups"*. Ibid.

Figure 8: Average odds difference for identifying gender biases

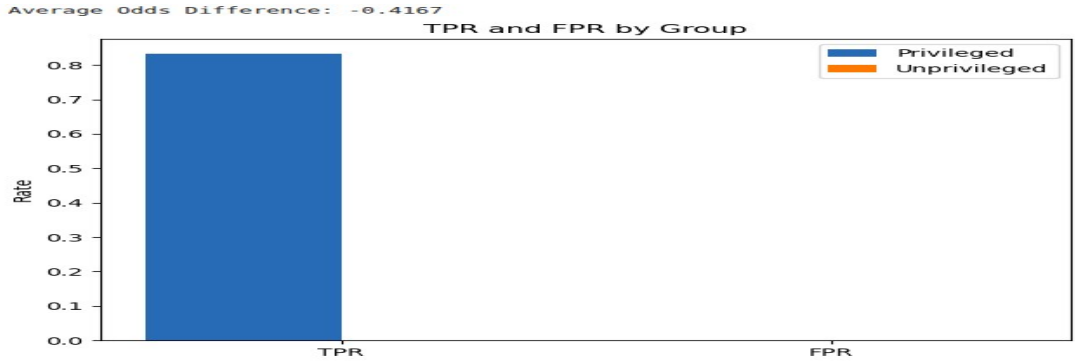

Once we have detected this amount of bias in the dataset, the next step is to calibrate it towards a more fair system. The disadvantages can be re-calibrated in percentages in order to equal their probability of being hired. For instance, let's calibrate the nationality.

Table 1: Percentages of natural persons' nationality within the dataset

Americans = min: 20%, most likely: 40%, max: 60%
Non-Americans = min: 50%, most likely: 70%, max: 90%

In the previous example, Americans are clearly benefited. We may assign to them a most likely vulnerability of 40% of vulnerability, while the vulnerability of other groups of non-Americans can be calibrated at a most likely of 70%. Then the input shall play a role as a decision factor in a model ontology. A good option is designing an ontology (based on the FAIR model ontology) with a *Biased Ranking Algorithm* risk scenario, where the vulnerability factor can be calibrated in order to benefit the conditions of specific disadvantaged groups of natural persons.

Figure 9: A risk scenario of Biased Ranking Algorithms from an AI users' perspective

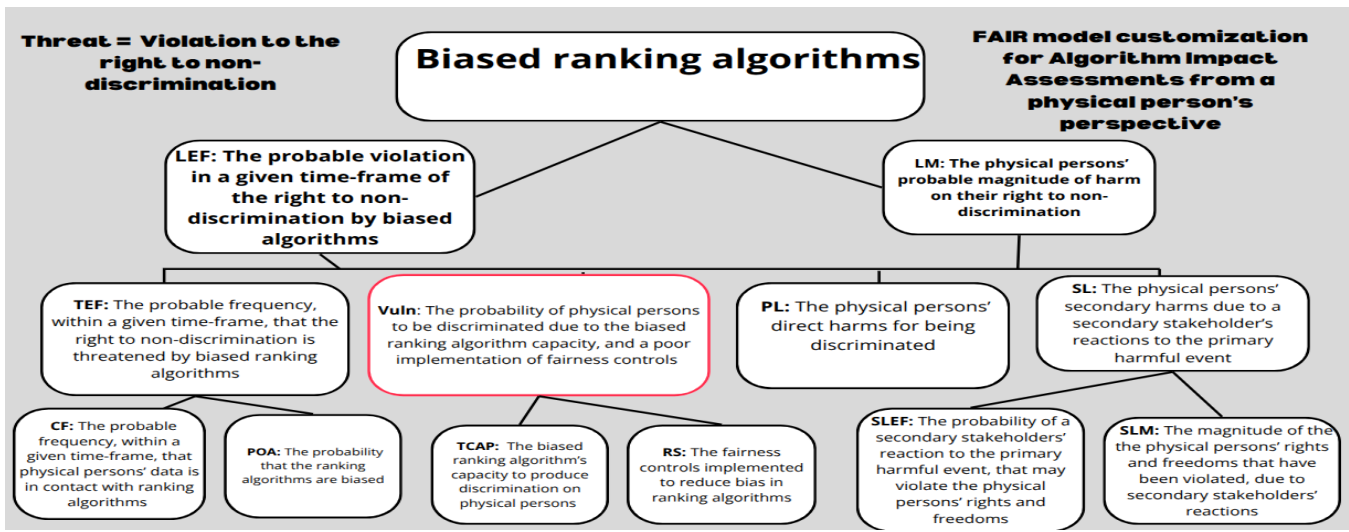

Here we can change the vulnerability input values for non-Americans. From here, we could apply a deterministic procedure such as the PERT formula (a+4b+c / 6)[36] in order to assign the minimum, most likely, and maximum values. We can also apply a stochastic procedure to the calibrated intervals by using the Monte Carlo simulation[37], with the goal of getting an accurate prediction interval. Nevertheless, a compliance focused model ontology from a AI deployers perspective may be based on a Fairness Value at Risk (F-VaR), as National Competent AI Authorities will have to estimate the amount of damage to the fundamental rights of natural persons in each fairness violation. The following example shows an AI risk ontology[38] to model that may help calibrating an AI risk analysis from a compliance risk perspective of AI deployers:

Figure 10: A fairness risk scenario concerning AI Act's compliance

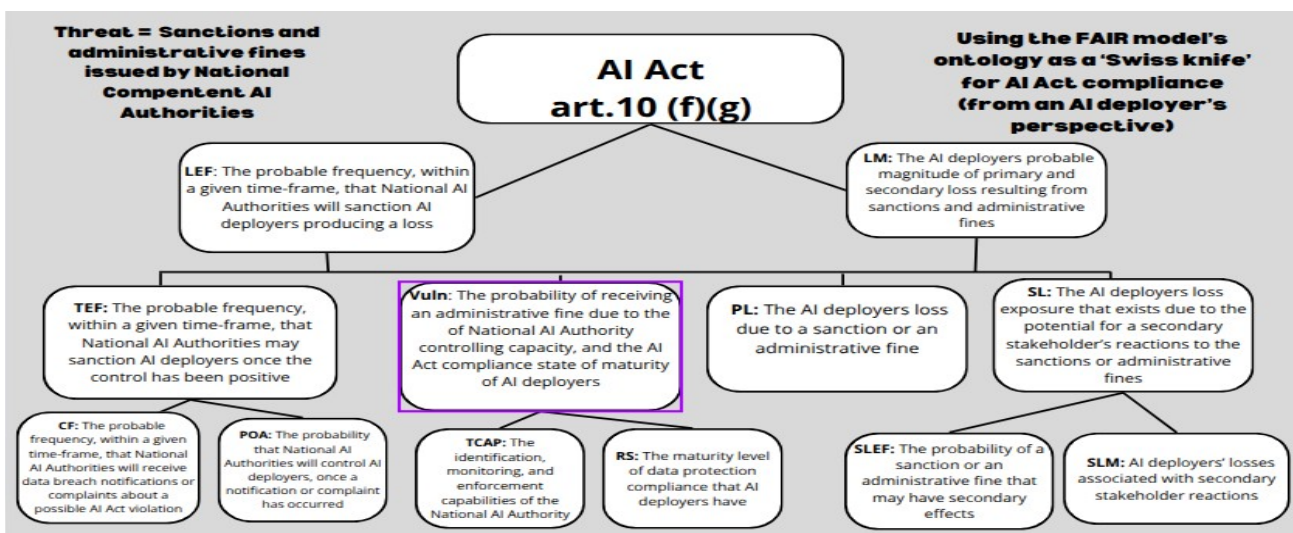

## 3. Modeling operational risk: accuracy and robustness

In the context of the AI Act, accuracy and robustness are defined in the AI Act as *"relevant performance metrics"[39]*. Accuracy is understood as the acceptable prediction intervals of an AI system. Robustness is understood as safety performance measures. Accuracy and robustness error metrics are crucial in order *"to measure AI prediction capabilities"*[40]. There are several metrics used in classification and regression problems. Classification metrics are usually based on the number of true positives, false positives, true negatives, and false negatives. For instance, well known metrics for classification are: accuracy[41], precision[42], recall[43]. Common regression metrics are Root Mean Squared Error (RMSE)[44] or Mean Absolute Error (MAE)[45]. While LLMs use next-token prediction metrics[46], LLMs are trained with classification-style objectives, and regression losses are useful in fine-tuning or auxiliary tasks[47]. This means that classification and regression metrics are fundamental for simple predictive models, and complex AI systems. For instance, let's consider an adversarial machine learning scenario of data poisoning that can lead to AI backdoor hallucinations. The threat community could be a group of cybercriminals with access to the AI system who can poison the training data of patients with a probability of cancer. The vulnerability could be the lack of access controls of the IA deployer. The consequence of the system's AI users may be the probable wrong answers that produce financial and psychological impact on potential cancer patients. The consequence of the AI provider and the AI provider may be the financial losses.

---

[36] "The characteristic of this technique is basing on a beta distribution and an optimistic, pessimistic, and most likely assessment". Milan Mirković, "Triangular Distribution and PERT Method vs. Payoff Matrix for Decision-Making Support in Risk Analysis of Construction Binding: A Case Study", *Architecture and Civil Engineering* 18, (2020): 290.
[37] See, Jack Freund and Jack Jones, *Measuring and Managing Information Risk: A FAIR Approach* (United States: Elsevier, 2015), 101.
[38] See, Luis Enríquez, "Using the Fair Model as Swiss Army Knife of Privacy Uncertainty Quantification for GDPR", FAIR Institute, Washington, 2024, https://www.fairinstitute.org/blog/fair-model-privacy-uncertainty-quantification-gdpr.
[39] Regulation (EU) 2024/1689 of the European Parliament and of the Council of 13 June 2024, *op. cit*., article 15(2).

**3.1. Classification.** The following synthetic dataset includes ten patients, where the features are *age, genetics, habits,* and *has_cancer* as the classification response, where the risk acceptance threshold is 50%. This means that patients below the threshold are labeled with '0' or no cancer, and patients with more than 50% are labeled as patients with a considerable probability of getting cancer.

Figure 12: A dataset of 10 natural persons with 3 features and the probability of getting cancer

|   | age | genetics | habits | has_cancer |
|---|-----|----------|--------|------------|
| 0 | 25  | 0.1      | 0.2    | 0          |
| 1 | 30  | 0.3      | 0.1    | 1          |
| 2 | 35  | 0.2      | 0.3    | 0          |
| 3 | 40  | 0.4      | 0.5    | 1          |
| 4 | 45  | 0.6      | 0.4    | 1          |
| 5 | 50  | 0.7      | 0.6    | 1          |
| 6 | 55  | 0.5      | 0.3    | 0          |
| 7 | 60  | 0.9      | 0.6    | 1          |
| 8 | 65  | 0.8      | 0.9    | 1          |
| 9 | 70  | 1.0      | 1.0    | 1          |

---

[40] Luciano Floridi, Matthias Holweg, *et al.,* capAI, A procedure for conducting conformity assessment of AI systems in line with the EU Artificial Intelligence Act, *op.cit.*, 35.
[41] *"Used in classification tasks where an approximately equal number of samples belong to each class". Ibid.*
[42] *"Used when the cost of False Positive is high". Ibid.*
[43] *"Used when the cost of False Negative is high". Ibid.*
[44] *"The RMSE is more appropriate to represent model performance than the MAE when the error distribution is expected to be Gaussian".* Tianfeng Chai and Roland Draxler, "Root mean square error (RMSE) or mean absolute error(MAE)? – Arguments against avoiding RMSE in the literature"*, Geoscientific Model Development 7,* no. 3 (2014):1247.
[45] *"The mean absolute error (MAE) is another useful measure widely used in model evaluations". Ibid.*
[46] Long Ouyang, Jeff Wu, *et al.,* "Training language models to follow instructions with human feedback", ArXiv: 2203.02155v1, 2022, 2.
[47] *Ibid.,* 3.

Figure 13: The predictions for the 10 natural persons

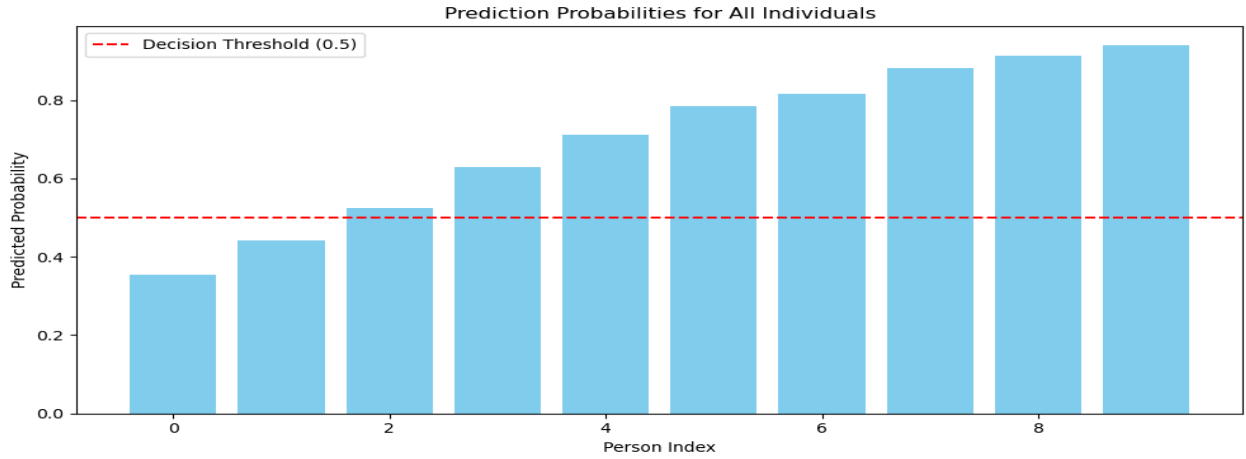

The logistic regression model can provide the probability of each patient, despite the fact of being classified in two groups. For the sake of showing classification metrics, accuracy and recall will be plotted:

Figure 14: Accuracy metrics

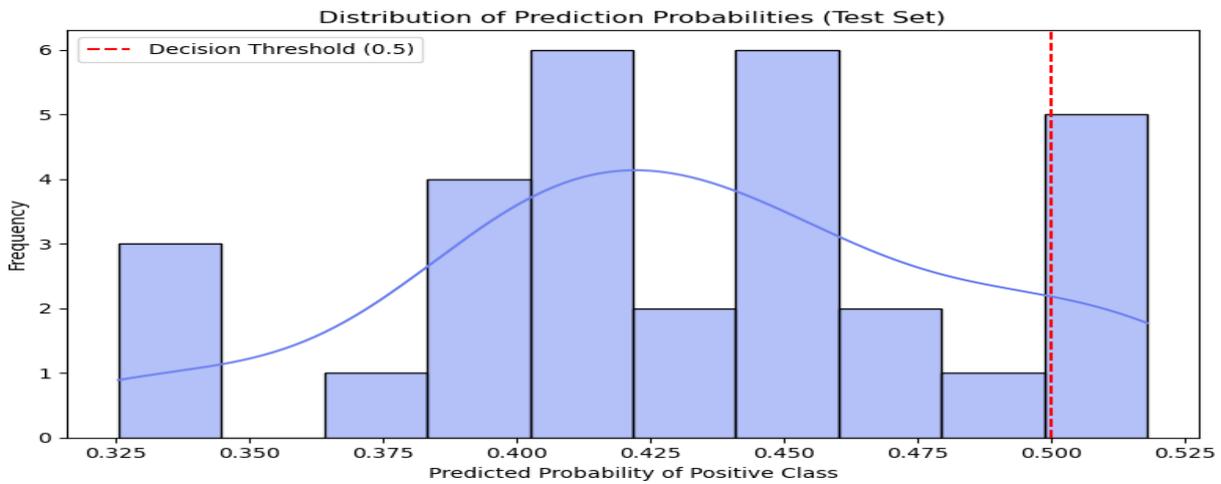

Figure 15: Recall metrics

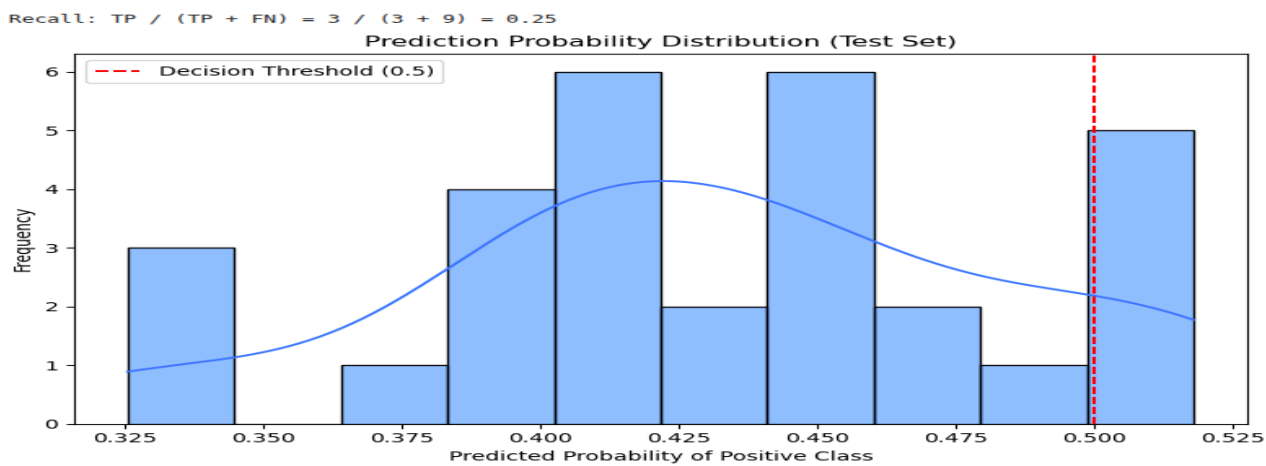

**3.2. Regression.** If we consider this hypothetical example as accurate, it is feasible to use the same example in an adversarial machine learning risk scenario of data poisoning with regression metrics. The historical dataset includes the average of data poisoning losses in the previous year:

Figure 16: A dataset of 10 administrative fines

|   | productivity_loss | reputation_damage | administrative_fine | breach_cost |
|---|---|---|---|---|
| 0 | 20000 | 10000 | 4000 | 52000 |
| 1 | 22000 | 12000 | 4500 | 73000 |
| 2 | 24000 | 9000 | 5000 | 76000 |
| 3 | 28000 | 20000 | 8000 | 119000 |
| 4 | 30000 | 25000 | 8500 | 151000 |
| 5 | 33000 | 30000 | 9000 | 181000 |
| 6 | 35000 | 28000 | 9500 | 203000 |
| 7 | 40000 | 35000 | 10000 | 237000 |
| 8 | 45000 | 40000 | 11000 | 283000 |
| 9 | 50000 | 45000 | 12000 | 318000 |

Applying a linear regression model with the RMSE metrics can provide good information about the predicted breach cost vs. the actual breach cost:

Figure 17: Comparison between actual and predicted administrative fines

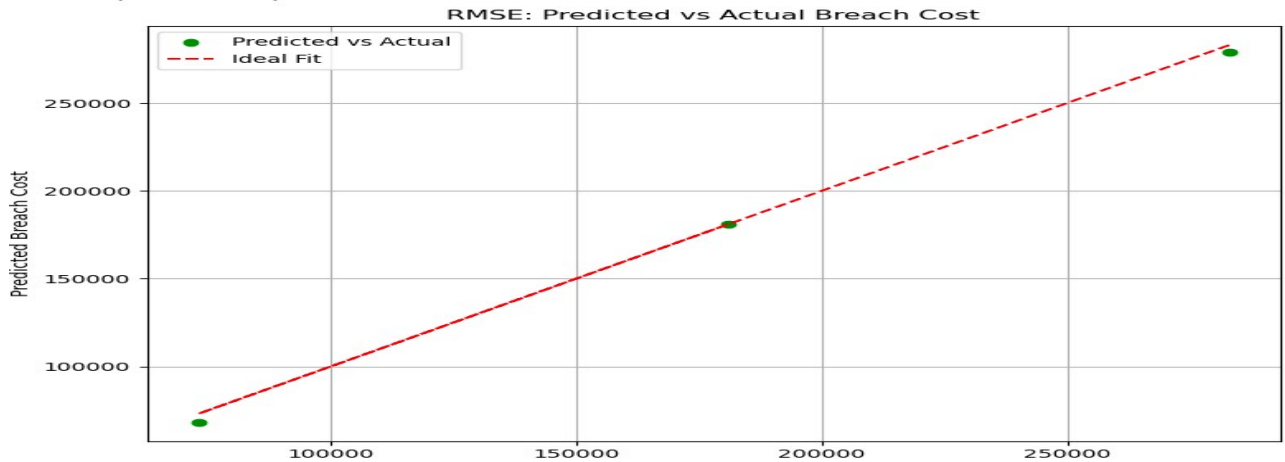

An accuracy & robustness Value at Risk (AR-VaR) may also be implemented when the risk scenario is linked with consumer protection or other legal areas. The ontology model would be similar to the ones already presented in the personal data and fairness risk ontologies.

## 4. Modeling Information security: Integrating personal data protection, fairness, accuracy and robustness

Once we get the output of a data protection, fairness, accuracy and robustness assessments, it is very convenient to integrate them into a single operational information security risk model. The FAIR model is ideal for information security risk scenarios related to artificial intelligence such as AI hallucinations[48], and adversarial machine learning[49]. The trick is integrating personal data protection, accuracy, robustness, and fairness in the right factors, while decomposing the risk scenario problem.

**4.1. Personal data.** It shall be calculated in its own risk ontology. The rationale of a personal data value at risk is based on an intelligence assessment about the sanctioning psychology of the data protection authority, and the maturity level of data protection compliance (such as GDPR compliance). The logic of the Jurimetrical Pd-VaR[50] relies on solving the difficulty of calibrating the material impact of a data protection risk on the rights and freedoms of the data subjects, which is then complemented with the calibrated Pd-VaR[51]. Instead of blindly guessing such impact, a very useful alternative is understanding the sanctioning psychology of the Data Protection Authority by using qualitative and quantitative information and argument retrieval analytical methods and adding the results as a secondary loss within an information security risk scenario linked to AI.

---

[48] *"To date, a precise and universally accepted definition of "hallucination" remains absent in the discussions related to this in the increasingly broader field of AI"*. However, Hallucination *"refers to instances where non-existent objects are erroneously detected or incorrectly localized"*. Negar Maleki, Balaji Padmanabhan, *et al.*, "AI Hallucinations: A Misnomer Worth Clarifying", arXiv:2401.06796v1, 2024, 1.

[49] See, Andrew McCarthy, Essam Ghadafi, *et al.*, "Defending against adversarial machine learning attacks using hierarchical learning: A case study on network traffic attack classification", *Journal of Information Security and Applications 72*, (2023): 2.

**4.2. Fairness.** It can also be calculated in an independent model ontology, and the output of the loss can be integrated as a secondary loss within the FAIR model. In the near future, we may also get fairness-related data coming from existing administrative fines issued by National AI authorities. Quantitative information and argument retrieval analytical methods may also be applied in order to calibrate a Fairness Value at Risk (F-VaR). The logic behind it relies on the difficulty of calibrating the loss of AI users, following the logic behind a Personal Data Value at Risk previously explained.

**4.3. Accuracy and robustness.** They may be integrated as resilience strength because their purpose is the good function and redundancy of the system that will mitigate the risk of harm against the fundamental rights of natural persons. These factors shall be weighed, combining them with the resilience level of the risk controls that shall provide protection from a threat community trying to compromise the system. In a nutshell, a bad accuracy and robustness assessment within the model would result in harm to the AI users, just like a threat community may exploit vulnerabilities of confidentiality, integrity, and availability controls. Therefore, in consumer protection legal risk scenarios, they may also become part of the *fines and judgments'* secondary losses factor.

**4.4. Information security.** It may be convenient in many risk scenarios to combine the data protection, fairness, accuracy and robustness rationales into an information security model. For instance, let's consider data poisoning combined with a biased algorithm ranking risk scenario. The protected asset is confidential data in CSV format, stored in a RAG environment. The threat community are cybercriminals. The purpose of the attack is changing the weights of a decision-making process in order to favor men over women. Despite that this risk scenario is about fairness, it has the four risk perspectives:

---

[50] See, Luis Enriquez, Personal data breaches : towards a deep integration between information security risks and GDPR compliance risks, *op. cit.*, 225.
[51] *Ibid.*

Table 2: A description of each AI risk regulatory dimension

- Data protection risk: Confidential data of natural persons will be accessed by the attacker without authorization (GDPR, articles 5(f), 32).
- Fairness risk: The selection processed will be biased (AI Act, article 10 (f) (g)).
- Accuracy and robustness risk: Even though that the classification metrics are properly applied, the result of the selection will be wrong because data was contaminated (AI Act, article 15 (1) (2) (3)).
- Information security risk: A security policy has been violated, as access controls failed. (AI Act, article 15 (4) (5)).

**5. Artificial Intelligence Value at Risk (AI-VaR).** These AI risk dimensions can be included in a classic FAIR model, and the predicted interval will provide the input of an AI-VaR. To make it simple, it will only be included Productivity as a Primary Loss, and Fines and Judgments as a Secondary Loss, with the average of a Personal Data Value at Risk and a Fairness Value at Risk. The Resistance Strength factor will have two sources: information security controls that may fail because the access controls are not good enough, while the accuracy and robustness of the AI system may fail due to the lack of model performance checking controls. In this example, they will be included with the following input values:

Table 3: Input values

Threat Event frequency: Min = 1; Most Likely = 5; Max = 10.
Threat Capability: Min = 20%; Most Likely = 50%; Max = 80%.
Resistance Strength: Min: 10%; Most Likely = 20%; Max = 30%.
Primary Loss (Productivity): Min = $1000; Most Likely = $4000; Max = $8000.
Secondary Loss Frequency (Fines and Judgments): Min= 60%; Most Likely = 80%; Max= 100%.
Secondary Loss Magnitude: Min = $4000; Most Likely = $8000; Max = $14000.

Figure 18: A risk scenario of data poisoning and Biased Ranking Algorithm (classic FAIR model)

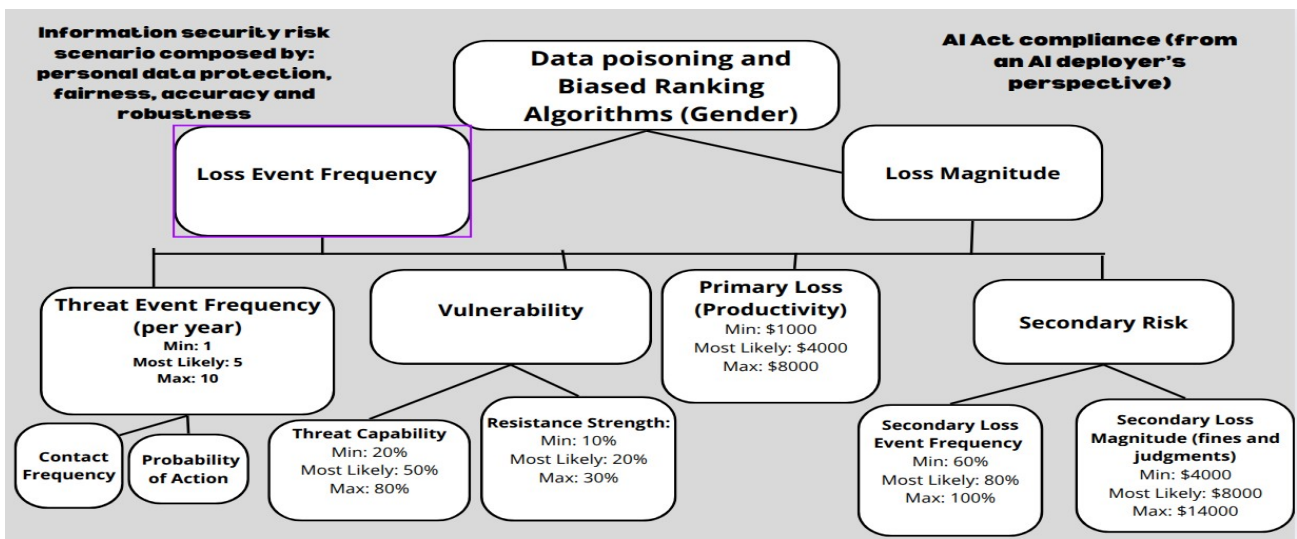

From this model, we can implement a Monte Carlo analysis with 1000 simulations. It is possible to ping the distribution at certain percentiles with the aim of setting limits. In the current Monte Carlo simulation, the Annual Loss Expectancy is $55 884. The P10th is $30 743. The P90th is $87 879.

Figure 19: A Monte Carlo analysis and raw distribution

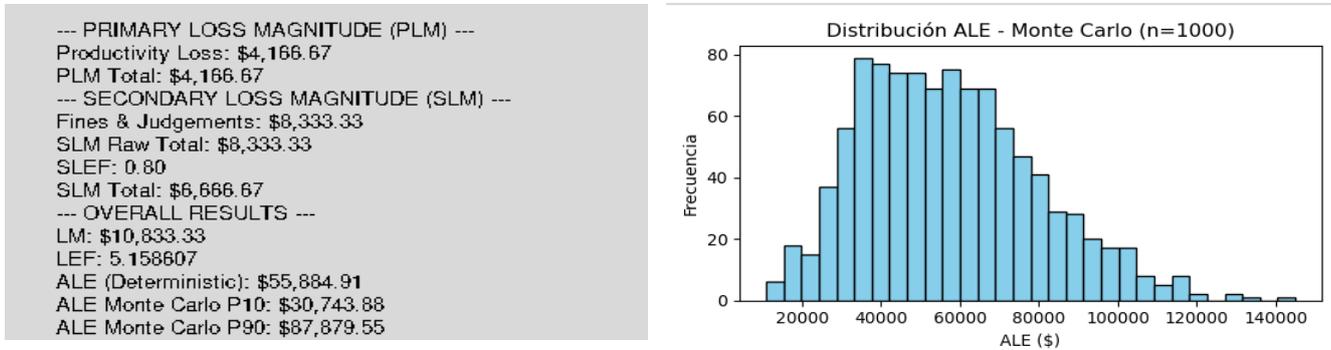

For obtaining an AI-VaR, it is necessary to set an interval. Unfortunately, the previous raw distribution is unstable around the low and high limits. A convenient way to implement the VaR in these cases is the an adjusted version of a Value at Risk[52] with truncated quantiles, as it allows to choose an interval by setting a low and a high percentile. In the following example, the lowest limit has been set at the 10$^{th}$ percentile, and the highest limit at the 90$^{th}$ percentile of the raw distribution. The result shows the worst annual loss of $81 589, at the 95$^{th}$ confidence level within a chosen interval between the p10th and the p90th.

Figure 20: An AI-VaR example

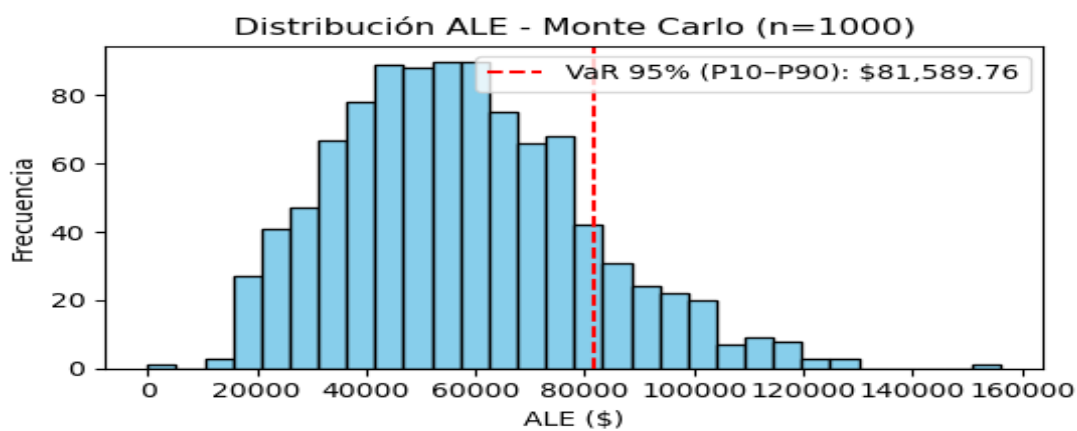

---

[52] It may be called a Conditional value at Risk, but the CvaR approach has been designed only for the worst losses (such as the tail's 5%). See, Tyrrel Rockafellar, Stanislav Uryasev, "Conditional value-at-risk for general loss distributions", *Journal of Banking and Finance,* Volume 26, Issue 7, (2002): 1443–1471.

# 6. Conclusion

This paper has explored the creation of meaningful metrics and adequate models for AI risk scenarios from several risk perspectives: personal data protection, fairness, accuracy, robustness and information security. Personal data protection metrics are fundamental to comply with frameworks such as the GDPR in the field of AI systems, focusing the analysis in two controversial spots: personal data as an input, and personal data as an output. Fairness metrics are essential to reduce bias conditions within the datasets, where the vulnerability factor in the model ontology helps to compensate the disadvantaged AI users. Accuracy and robustness metrics are essential for the right performance of an AI system, as a bad AI system performance may also threaten the fundamental rights of natural persons. Finally, information security scenarios have been presented as the holy grail for AI risk integration, as without information security, AI systems will certainly fail.

The integration of these four kinds of AI metrics in an information security risk scenario becomes a very challenging but feasible task. The goal of this approach is to first solve the data protection and fairness values at risk, as secondary losses, by using legal predictive analytics in order to understand the sanctioning psychology of the AI National Competent Authorities. Then, proceed to assess the accuracy and robustness assessments in order to combine them in a holistic operational risk scenario, where the loss due to the bad performance of the AI system shall be integrated as resistance strength and sometimes as a potential secondary loss. Yet, in several cases, it would be necessary to model data protection, accuracy, robustness, and fairness in their own risk ontologies in order to export their output into an information security risk scenario.